\documentclass[pre,showpacs,preprint]{revtex4}

\usepackage{amsmath}

\usepackage{graphicx}
\usepackage{dcolumn}
\usepackage{bm}

\begin{document}

\title{Vibrated granular gas confined by a piston}
\author{J. Javier Brey and M.J. Ruiz-Montero}
\affiliation{F\'{\i}sica Te\'{o}rica, Universidad de Sevilla,
Apartado de Correos 1065, E-41080, Sevilla, Spain}
\date{\today }

\begin{abstract}
The steady state of a vibrated granular gas confined by a movable
piston on the top is discussed. Particular attention is given to the
hydrodynamic boundary conditions to be used when solving the
inelastic Navier-Stokes equations. The relevance of an exact general
condition relating the grain fluxes approaching and moving away from
each of the walls is emphasized. It is shown how it can be used to
get a consistent hydrodynamic description of the boundaries. The
obtained expressions for the fields do not contain any undetermined
parameter. Comparison of the theoretical predictions with molecular
dynamics simulation results is carried out, and a good agreement is
observed for low density and not too large inelasticity. A practical way of introducing small finite density corrections to the dilute limit theory is proposed, to improve the
accuracy of the theory.

\end{abstract}


\maketitle

\section{Introduction}
\label{s1} The behavior of fluidized granular systems resembles in
many cases that of ordinary molecular fluids. Actually, it is by now
well established that generalized Navier-Stokes equations describe
quite accurately many of the experimental and numerical features of
granular flows, especially at low density and small inelasticity
\cite{Ha83,Ca90,Go03}. The justification for this fluid-like
description, and derivation of theoretical predictions for the
transport coefficients appearing in the equations, have been
intensively studied for some time. Quite often, idealized systems of
inelastic hard spheres or disks have been considered. For
mono-disperse  models of this kind, the studies carried out include
the derivation of the hydrodynamic equations to Navier-Stokes order
by using kinetic theory methods
\cite{LSJyCh84,JyR85,SyG98,BDKyS98,GyD99,ByD05}, direct Monte Carlo
simulation of (inelastic) kinetic equations \cite{BRyC96,ByC01},
molecular dynamics simulations \cite{He95}, and, very recently,
linear response theory \cite{DByB08,BDyB08}.

Interactions between grains are inherently inelastic. As a
consequence, the kinetic energy of isolated granular systems
decreases monotonically in time and, in order to keep them
fluidized, it is necessary to continuously supply energy to them. A
prototypical way of doing it is by vibrating one of the walls of the
container, usually the one at the bottom. Also often, the interest
focuses on the bulk properties of the system, i.e. on the behavior
of the system far enough from the walls, where the governing laws
are expected to be independent of the details of the boundaries.
Then, the most appropriate possible way of vibration for this
purpose is chosen. In many situations of interest, the criterium for
this choice is twofold: simplicity and avoiding undesired effects
such as the induction of propagating waves into the system. These
goals are formally achieved in the limits of very high frequency and
very small amplitude; the former as compared with the typical
relaxation frequency of the granular fluid next to the wall, and the
latter with its mean free path. Additional simplifications occur if
collisions of the grains with the wall are taken as elastic and if
the wall is assumed to move with a sawtooth profile
\cite{McyB97,McyL98}.

A kind of idealized walls often used in kinetic theory and particle
simulations are the so-called thermal walls. By definition, the flux
of after-collision particles leaving it corresponds to a Maxwellian
flux with the temperature parameter characterizing the wall
\cite{DyvB57}. Therefore, the shape of the velocity distribution of
the particles moving away from the thermal wall is independent of
the distribution of the ingoing particles. It is evident that
thermal walls also provide a mechanism to compensate the energy
dissipated in inelastic collisions, once the granular gas in the
vicinity of the wall tends to have a temperature smaller than the
one of the wall. It is then not surprising that thermal walls have
also been extensively used in the literature of fluidized granular
gases since a decade ago \cite{GZyB97,ByC98}. Nevertheless, it must
be stressed that it is not at all evident that thermal walls
correspond to any limit of a rapidly vibrating plate. Actually, it
has been shown that in some cases the stability of systems driven by
a thermal wall and by a wall vibrating in a sawtooth way can differ
\cite{BRMyG02}.

The granular system considered in this paper is confined between two
parallel plates in presence of a gravity force. The mission of the
one at the bottom of the system is to fluidize the granular medium,
as in the previous studies mentioned above. On the other hand, the
wall on the top is floating, in the sense that it can move in the
vertical direction, being supported by the granular fluid below it.
As a consequence, the position and motion of the upper wall is
interrelated with the state of the granular media below it, and the
boundary conditions to be imposed to the hydrodynamic equations,
following from the interaction between the gas and the piston, must
be determined in a self-consistent way. Here, the steady state
eventually reached by the granular gas between the two plates  will
be investigated, using the hydrodynamic description provided by the
Navier-Stokes equations for inelastic hard spheres. Collisions of
the grains with the piston will be modeled as hard inelastic
collisions. This defines in a deterministic way the mechanical
interaction between the particles and the movable wall. The question
addressed afterwards is how to translate it into an appropriate
boundary condition to be used in the context of hydrodynamics. Here
this will be done by means of an intermediate stage in which an
exact boundary condition for the kinetic theory description of the
system is formulated. This condition relates the distribution
functions of particles leaving the wall and approaching it, by
expressing the conservation of the particles flux at the piston.

An additional boundary conditions is required to determine the
hydrodynamic fields. It can be obtained, in equivalent ways, from
the energy flux at the vibrating bottom wall or from the global
balance of energy in the system. In this way, the theoretical
prediction is completed, and explicit expressions for the fields
with no adjustable parameters are derived. In order to verify the
accuracy of this description, the predictions are compared with
molecular dynamics simulation results. As in other steady states of
granular gases, the range of applicability of the theory is
restricted to values of the restitution coefficient for gas
particles collisions close to unity, because of the coupling between
inelasticity and gradients. Under these conditions, reasonable
agreement between theory and simulations is observed in the bulk of
the system, i.e.  outside the kinetic boundary layers next to  the
walls. This confirms the validity of the hydrodynamic description,
including the needed boundary conditions, to describe vibrated
granular gases in quite realistic situations.

The plan of the paper is as follows. In Sec.\ \ref{s2}, a previously
derived \cite{BRyM01} stationary solution of the inelastic
Navier-Stokes equations for a vibrated dilute granular system in
presence of gravity is reviewed.  The main results are explicit
expressions for the hydrodynamic fields of the system having two
arbitrary parameters. Additionally, they involve the height of the
system. The results are particularized in Sec. \ref{s3} for the
granular gas between two plates described above. Also, the boundary
effects following from the interaction between the gas and the
piston on the top are formulated as a condition for the gas-piston
distribution function at contact. In the same section and in
Appendix \ref{ap2}, it is discussed why this is the appropriate
starting point to derive the hydrodynamic boundary condition and to
introduce self-consistent approximations, like those used in Sec.
\ref{s4}. An analysis along the same lines of the hydrodynamic
boundary effects due to the vibrating wall at the bottom is
presented in Appendix \ref{ap3}. Also in Sec.\ \ref{s4}, the derived
boundary conditions are used to identify the arbitrary constants in
the hydrodynamic profiles derived in Sec. \ref{s2}. There are no
adjustable parameters in these expressions.

The comparison of the obtained theoretical predictions with
molecular dynamics simulation results is carried out in Sec.
\ref{s5} for two-dimensional systems. It includes both the detailed
description provided by the hydrodynamic profiles and also some
global properties, like the average position of the piston and the
the balance of the total energy of the system. A fairly good
agreement is observed, especially if some (small) finite density
effects are partially incorporated into the hydrodynamic
description, through the equation of state of the gas. The papers
ends with a short summary and some general comments.

\section{The general one-dimensional solution}
\label{s2} In this section, some of the results already discussed in
ref. \cite{BRyM01} will be shortly reviewed and summarized for the
sake of completeness. The system considered is a dilute granular gas
composed of $N$ equal smooth inelastic hard spheres ($d=3$) or disks
($d=2$) of mass $m$ and diameter $\sigma$. The position and velocity
of grain $i$ will be denoted by ${\bm r}_{i}$ and ${\bm v}_{i}$,
respectively. The inelasticity of collisions between grains is
modeled by means of a constant, velocity independent, coefficient of
normal restitution $\alpha$, defined in the interval $ 0 < \alpha
\leq 1$. There is an external gravitational field acting on the
system, so that each particle is submitted to a force $-m g_{0}
\widehat{\bm e}_{z}$, where $g_{0}$ is a positive constant and
$\widehat{\bm e}_{z}$ is the unit vector in the positive direction
of the $z$ axis.

For steady states with vanishing macroscopic flow and gradients only
in the direction of the external field, i.e. the $z$ axis, the
inelastic hydrodynamic Navier-Stokes equations of this system reduce
to \cite{BRyM01}
\begin{equation}
\label{2.1} \frac{\partial p}{\partial z} =-nmg_{0},
\end{equation}
\begin{equation}
\label{2.2} \frac{2}{nd} \frac{\partial}{\partial z} \left( \kappa
\frac{\partial T}{\partial z} + \mu \frac{\partial n}{\partial z}
\right) - T \zeta^{(0)}=0.
\end{equation}
Here, $n(z)$ is the local number of particles density, $T(z)$ the
local granular temperature, and $p(z)=n(z)T(z)$ the  pressure. The
temperature is defined from the kinetic energy in the usual way, but
with the Boltzmann constant set equal to unity. Moreover, $\kappa $
is the thermal heat conductivity and $\mu $ the diffusive heat
conductivity, that is peculiar of granular systems. More
specifically, the generalized Fourier law giving the heat flux
$q_{z}$ in the system is
\begin{equation}
\label{2.2a} q_{z}= - \kappa \frac{\partial T}{\partial z} - \mu
\frac{\partial n}{\partial z}.
\end{equation}
Finally, $\zeta^{(0)}(z)$ is the cooling rate accounting  for the
energy dissipated in collisions. Upon deriving Eq.\ (\ref{2.1}), use
has been made of the local equation $p(z)=n(z)T(z)$, where $p$ is
the hydrodynamic pressure, valid in the low density limit.

The expressions of the transport coefficients and the cooling rate
appearing in the above expressions can be written in the form
\cite{BDKyS98,ByC01}
\begin{equation}
\label{2.3} \kappa(\alpha,T) = \kappa^{*}(\alpha) \kappa_{0}(T),
\end{equation}
\begin{equation}
\label{2.4} \mu (\alpha,T) = \mu^{*}(\alpha) \mu_{0}(T),
\end{equation}
\begin{equation}
\label{2.5} \zeta^{(0)}(\alpha,T) = \zeta^{*} (\alpha)
\frac{p}{\eta_{0}(T)}\, ,
\end{equation}
with $\kappa_{0}(T)$ and $\eta_{0}(T)$ being the elastic ($\alpha =
1$) values of the thermal heat conductivity and the shear viscosity,
\begin{equation}
\label{2.6} \kappa_{0}(T) = \frac{d(d+2)^{2} \Gamma  \left( d/2
\right) }{16 (d-1) \pi^{\frac{d-1}{2}}} \left( \frac{T}{m}
\right)^{1/2} \sigma^{-(d-1)},
\end{equation}
\begin{equation}
\label{2.7} \eta_{0}(T)= \frac{(d+2) \Gamma \left(d/2 \right)}{8
\pi^{\frac{d-1}{2}}} \left( mT \right)^{1/2} \sigma^{-(d-1)},
\end{equation}
respectively, and
\begin{equation}
\label{2.8} \mu_{0}(T) = \frac{T \kappa_{0}(T)}{n}\, .
\end{equation}
The dimensionless quantities $\kappa^{*}$, $\mu^{*}$, and
$\zeta^{*}$ are given in appendix \ref{ap1}. They only depend on
the restitution coefficient $\alpha$.

Equation (\ref{2.2}) shows the physically evident feature that, in
this state,  hydrodynamic gradients are induced by the inelasticity,
through the cooling rate. Consequently, a restriction to small
gradients, as it is the case in the Navier-Stokes approximation used
above, also implies a limitation on the value of $\alpha$ for which
the theory can be expected to apply. The interval of values of this
parameter for which the theory actually provides an accurate
description is very hard to determine a priori.

The system is supposed to be confined between two parallel walls
located at $z=0$ and $z=L$, respectively. The nature of this two
walls will be specified and discussed later on. It is convenient to
introduce a dimensionless length scale $\xi$ by \cite{BRyM01}
\begin{equation}
\label{2.9} \xi= \sqrt{a(\alpha)} \left[ \int_{z}^{L} dz^{\prime}\,
\frac{1}{\lambda (z^{\prime})} + \frac{\sigma^{d-1}p_{L}}{m g_{0}}
\right],
\end{equation}
where
\begin{equation}
\label{2.10} \lambda (z) \equiv \left[ \sigma^{d-1} n(z)
\right]^{-1}
\end{equation}
is proportional to the local mean free path, $p_{L} \equiv p(z=L)$
is the pressure of the gas next to the wall located at $z=L$, and
\begin{equation}
\label{2.11} a(\alpha) \equiv \frac{ 32 (d-1) \pi^{d-1}
\zeta^{*}(\alpha)}{(d+2)^{3}  \Gamma^{2}\left( d/2 \right) \left[
\kappa^{*}(\alpha)-\mu^{*} (\alpha) \right]}\, .
\end{equation}
The $\xi$ coordinate is a monotonic decreasing function of $z$,
varying between
\begin{equation}
\label{2.12} \xi_{M} \equiv \xi (z=0) = \sqrt{a(\alpha)}
\sigma^{d-1} \left( N_{z}+\frac{p_{L}}{mg_{0}} \right)
\end{equation}
and
\begin{equation}
\label{2.13} \xi_{m}\equiv \xi(z=L) = \sqrt{a(\alpha)} \frac{
\sigma^{d-1} p_{L}}{m g_{0}}.
\end{equation}
In Eq.\ (\ref{2.12}), $N_{z}$ denotes the number of particles in the
system per unit of section $W$ (length or area) perpendicular to the
external field, $N_{z} \equiv N/W$. It must be noted that the
variation interval of $\xi$ depends on $p_{L}$ and $N_{z}$, but not
on the value of $L$.

The physical meaning of $\xi$ can be illustrated by realizing that
it is proportional to the local pressure at the corresponding height
$z$. This follows from Eq.\ (\ref{2.1}), that leads to
\begin{equation}
\label{2.13a} p(z) = m g_{0} \int_{z}^{L} d z^{\prime}\,
n(z^{\prime}) + p_{L}= \frac{m
g_{0}}{\sqrt{a(\alpha)}\sigma^{d-1}}\, \xi.
\end{equation}

By substituting Eq.\ (\ref{2.1}) into Eq.\ (\ref{2.2}) and doing the
change of variable defined in Eq.\ (\ref{2.9}), it is obtained:
\begin{equation}
\label{2.14} \xi \frac{\partial^{2} T^{1/2}}{\partial \xi^{2}} +
b(\alpha) \frac{\partial T^{1/2}}{\partial \xi} - \xi T^{1/2} =0,
\end{equation}
with
\begin{equation}
\label{2.15} b(\alpha) \equiv \frac{2 \kappa^{*}
-\mu^{*}}{2(\kappa^{*}-\mu^{*})}\, .
\end{equation}
The general solution of the above differential equation reads
\cite{AyS65}
\begin{equation}
\label{2.16} T^{1/2}(\xi) = A \xi^{-\nu} I_{\nu}(\xi)+B \xi^{-\nu}
K_{\nu} (\xi),
\end{equation}
where $A$ and $B$ are constants to be identified from the boundary
conditions,
\begin{equation}
\label{2.17} \nu (\alpha) \equiv \frac{b(\alpha)-1}{2} =
\frac{\mu^{*}}{4 \left( \kappa^{*}-\mu^{*} \right) }\, ,
\end{equation}
and $I_{\nu}$ and $K_{\nu}$ are the modified Bessel functions of
first and second kind, respectively \cite{AyS65}.

Therefore, for the system being considered, the pressure and
temperature profiles are given by Eqs.\ (\ref{2.13a}) and
(\ref{2.16}), respectively. Consequently, the density profile is
\begin{equation}
\label{2.18} n(\xi) = \frac{p(\xi)}{T(\xi)} = \frac{m g_{0}
\xi}{\sqrt{a(\alpha)} \sigma^{d-1} \left[ A\xi^{-\nu} I_{\nu}(\xi)
+B \xi^{-\nu} K_{\nu} (\xi) \right]^{2}}\, .
\end{equation}
Finally, the transformation from the $\xi$ coordinate to the $z$ one
is given by
\begin{equation}
\label{2.19} z= \frac{1}{\sqrt{a(\alpha)}\sigma^{d-1}}
\int_{\xi}^{\xi_{M}} \frac{d \xi^{\prime}}{n(\xi^{\prime})},
\end{equation}
that follows from Eq.\ (\ref{2.9}). It is worth to mention that the
presence of the diffusive heat conductivity $\mu$ in the above
expressions is not at all irrelevant. Predictions implied by its
existence have been checked both by particle simulations and
experimentally \cite{ByR04,CyR91,ByK03,WyH08}. To proceed any
further, the boundary conditions of the system at the top ($z=L$)
and the bottom ($z=0$) must be specified. This will be done in the
next section.

\section{Closed system with a piston. The kinetic boundary condition.}
\label{s3} In order to maintain the system fluidized, it will be
assumed that energy is being continuously supplied to it through the
wall located at the bottom, and that this is achieved by vibrating
it. The simplest possible way of vibration will be considered here,
namely with a sawtooth velocity profile. This means that all the
particles colliding with the wall find it with the same upwards
velocity $v_{W}$ \cite{McyB97,McyL98}. Moreover, the amplitude of
vibration of this wall is taken much smaller than the mean free path
of the grains in its vicinity. As a consequence, the position of the
wall can be taken in practice as fixed at $z=0$ with very good
accuracy. Finally, since the main reason to introduce this vibrating
wall is to keep the granular matter fluidized, collisions of
particles with it will be considered as elastic, for the sake of
simplicity. Of course, all the above corresponds  to a very idealized
wall that can not be fully implemented in actual experiments.

\begin{figure}
\includegraphics[scale=0.5,angle=0]{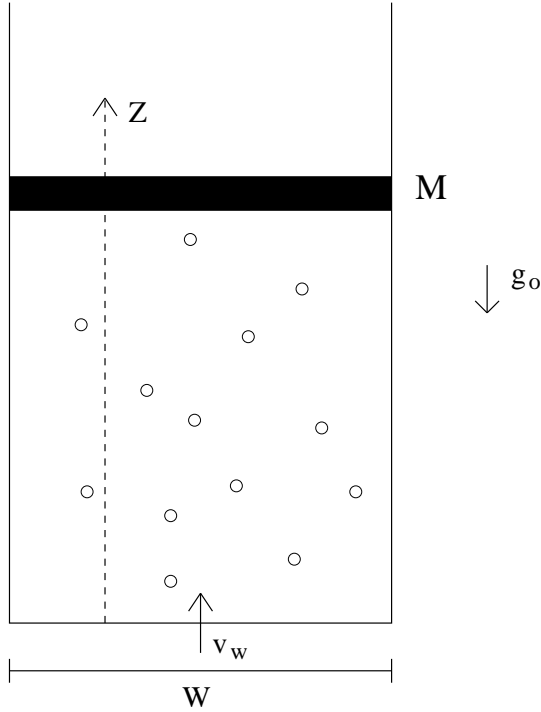}
\caption{Sketch of the system studied in this paper.\label{fig1}}
\end{figure}

Next, the upper boundary condition must be specified. The case of an
open system ($ L \rightarrow \infty$) was studied in ref.
\cite{BRyM01}. Here, a different physical situation will be
investigated. It will be considered that there is a movable lid or
piston on top of the gas, as illustrated in Fig.\ \ref{fig1}. The
piston has a finite mass $M$, contrary to the vibrating wall at the
bottom that is taken infinitely massive. The piston can only move in
the $z$ direction, remaining always perpendicular to it, i.e.
parallel to the bottom wall. Its position and velocity will be
denoted by $Z$ and $V_{z}$, respectively, so that $L$ corresponds to
the average value of $Z$ in the steady state. There is no friction
between the piston and the lateral walls of the container.

Collisions of particles with the piston on the top are smooth and
inelastic, with a velocity-independent coefficient of normal
restitution $\alpha_{P}$, $0<\alpha_{P} \leq 1$. Therefore, the
vector component, ${\bm v}_{\perp}$, of the velocity of the particle
perpendicular to the $z$ axis remains unchanged in a collision with
the piston,
\begin{equation}
\label{3.1} {\bm v}_{\perp}^{\prime}= {\bm v}_{\perp},
\end{equation}
where the prime is used here and henceforth to denote
after-collision quantities. On the other hand, when a particle with
a component $v_{z}$ of the velocity collides with the piston being
$V_{z}$ the velocity of the latter, these values change
instantaneously to
\begin{equation}
\label{3.2} v^{\prime}_{z} =v_{z} -\frac{M}{m+M}\, (1+\alpha_{P})
(v_{z}-V_{z}),
\end{equation}
\begin{equation}
\label{3.3} V^{\prime}_{z} =V_{z} +\frac{m}{m+M}\, (1+\alpha_{P})
(v_{z}-V_{z}).
\end{equation}
Therefore, in the collision the total momentum is conserved, the
relative velocity $g_{z} \equiv v_{z}-V_{z}$ changes to
$g_{z}^{\prime}=- \alpha_{P} g_{z}$, and there is a variation of the
total kinetic energy given by
\begin{equation}
\label{3.4} \Delta E = - \frac{mM}{2 (m+M)}\, (1-\alpha_{P}^{2})
g_{z}^{2}.
\end{equation}
For $\alpha_{P} <1$  there is a loss of energy in every collision
between a particle and the piston. The change in the $z$-component
of the momentum $P_{z}$ of the piston in a collision is
\begin{equation}
\label{3.5} \Delta P_{z} \equiv M(V^{\prime}_{z}-V_{z}) =
\frac{mM}{m+M}\, (1+\alpha_{P})g_{z}.
\end{equation}
Since a collision of a particle with the piston is only possible if
$v_{z}
> V_{z}$, it is $\Delta P_{z} >0$ for all the collisions, indicating
that momentum is continuously transferred from the gas to the
piston. A relevant relationship to be used in the following is
\begin{equation}
\label{3.6} d{\bm v}^{\prime\, } dV^{\prime}_{z} = \alpha_{P} d {\bm
v}\, dV_{z}.
\end{equation}
For later use, it is convenient to consider also the so-called
restituting collision corresponding to the velocities ${\bm v}$ and
$V_{z}$. It is defined by the velocities ${\bm v}^{*}$ and
$V_{z}^{*}$ leading as a consequence  of a collision to ${\bm v}$
and $V_{z}$. Their expressions are obtained directly by inverting
Eqs.\ (\ref{3.1})-(\ref{3.3}),
\begin{equation}
\label{3.7} {\bm v}^{*}_{\perp} = {\bm v}_{\perp},
\end{equation}
\begin{equation}
\label{3.8} v^{*}_{z} =v_{z} -\frac{M}{m+M}\,
\frac{1+\alpha_{P}}{\alpha_{P}} g_{z},
\end{equation}
\begin{equation}
\label{3.9}
 V^{*}_{z} =V_{z} +\frac{m}{m+M}\, \frac{1+\alpha_{P}}{\alpha_{P}}
g_{z}.
\end{equation}
Also, it is  $g^{*}_{z} = \alpha_{P}^{-1} g_{z}$ and
\begin{equation}
\label{3.10} d {\bm v}^{*} dV_{z}^{*} = \alpha_{P}^{-1} d {\bm v}
dV_{z}\, .
\end{equation}

The question now is how to translate the above collision rules into
one or more boundary conditions for the description of the granular
gas below the piston. To discuss this point in some detail, let us
introduce the two-body distribution function for the piston and the
gas, $\Phi ({\bm x}, Z, V_{z},t)$, defined for an arbitrary state as
\begin{equation}
\label{3.11} \Phi ({\bm x}, Z, V_{z},t)= N \int d{\bm x}_{2} \ldots
\int d{\bm x}_{N} \rho ({\bm x},{\bm x}_{2}, \ldots, {\bm
x}_{N},Z,V_{Z},t),
\end{equation}
where $\rho ({\bm x}_{1},{\bm x}_{2}, \ldots, {\bm
x}_{N},Z,V_{Z},t)$, with ${\bm x}_{i} \equiv \{ {\bm r}_{i},{\bm
v}_{i} \}$, is the distribution function for the system composed by
the piston and the $N$ grains at time $t$, and ${\bm x} \equiv \{
{\bm r},{\bm v} \}$. Therefore, $\Phi ({\bm x}, Z, V_{z},t)$ is
proportional to the probability density of finding the piston at
height $Z$ with velocity $V_{z}$ and a grain at position ${\bm r}$
with velocity ${\bm v}$, at time $t$. It is normalized as
\begin{equation}
\label{3.12} \int d{\bm x} \int_{0}^{\infty} dZ \int_{-
\infty}^{\infty} dV_{z}\, \Phi ({\bm x}, Z, V_{z},t)=N.
\end{equation}

The one-particle distribution function of the gas, $f({\bm x},t)$,
can be obtained from $\Phi$ by integration over the piston position
and velocity,
\begin{equation}
\label{3.13} f({\bm x},t) = \int_{0}^{\infty} dZ \int_{-
\infty}^{\infty} dV_{z}\, \Phi ({\bm x}, Z, V_{z},t).
\end{equation}
Similarly, the probability distribution for the piston,
$F(Z,V_{z},t)$, is given by
\begin{equation}
\label{3.14} F(Z,V_{z},t)= \frac{1}{N} \int d{\bm x}\, \Phi ({\bm
x}, Z, V_{z},t),
\end{equation}
and it is normalized to unity. No reference to any particular state
is involved in the above definitions.

For initial conditions in which all the particles are located below
the piston, the two-body distribution $\Phi$ at arbitrary later
times can be expressed in the form
\begin{equation}
\label{3.15} \Phi ({\bm x}, Z, V_{z},t) = \Theta (Z-z)\Phi_{0} ({\bm
x}, Z, V_{z},t),
\end{equation}
where $\Theta(x)$ is the Heaviside step function defined by $\Theta
(x) =1$ for $x \geq 0$ and $\Theta (x) =0$ for $x<0$. The function
$\Phi_{0} ({\bm x}, Z, V_{z},t)$ is not defined by Eq.\ (\ref{3.15})
for $z > Z$, and it can be considered as being regular everywhere as
well as its derivatives, without restriction. In order to discuss
the form of $\Phi_{0}$ when the position of the particle is taken
next to the piston, it is useful to decompose it in the form
\begin{equation}
\label{3.16} \Phi_{0} ({\bm x}, Z, V_{z},t)= \Phi_{+} ({\bm x}, Z,
V_{z},t) + \Phi_{-} ({\bm x}, Z, V_{z},t),
\end{equation}
with
\begin{equation}
\label{3.17} \Phi_{+} ({\bm x}, Z, V_{z},t) \equiv \Theta (g_{z})
\Phi_{0} ({\bm x}, Z, V_{z},t)
\end{equation}
and
\begin{equation}
\label{3.18} \Phi_{-} ({\bm x}, Z, V_{z},t)  \equiv \Theta (-g_{z})
\Phi_{0} ({\bm x}, Z, V_{z},t).
\end{equation}
In the last expression, it is $\Theta(-x) \equiv 1- \Theta (x)$.

Conservation of the flux of particles at the piston implies that
\begin{equation}
\label{3.19} \Phi_{+} ({\bm r}, {\bm v}^{*}, Z, V_{z}^{*},t) d {\bm
v}^{*}\, dV^{*}_{z}\, |g_{z}^{*}| \delta (Z-z) = \Phi_{-} ({\bm
r},{\bm v},Z,V_{z},t) d{\bm v}\, d V_{z}\, |g_{z}| \delta (Z-z),
\end{equation}
where ${\bm v}^{*}$ and $V_{z}^{*}$ are the restituting velocities
defined by Eqs.\ (\ref{3.7})-(\ref{3.9}). This is an exact
relationship, valid for arbitrary density, that can be derived also
by starting from the evolution equation for $\Phi ({\bm x}, Z,
V_{z},t)$ following from the Liouville equation for $\rho ({\bm
x}_{1},{\bm x}_{2}, \ldots, {\bm x}_{N},Z,V_{Z},t)$. Equation
(\ref{3.19}) is obtained by isolating the singular terms at $z=Z$
\cite{Lu99}. When Eqs. (\ref{3.7})-(\ref{3.10}) are employed into
Eq. (\ref{3.19}), it can be reduced to
\begin{equation}
\label{3.20} \Phi_{-} ({\bm r},{\bm v},Z,V_{z},t) \delta (Z-z)=
\alpha_{P}^{-2} \Phi_{+} ({\bm r},{\bm v}^{*},Z,V_{z}^{*},t)
\delta(Z-z).
\end{equation}
Combination of Eqs.\ (\ref{3.16}) and (\ref{3.20}) yields
\begin{equation}
\label{3.21} \Phi_{0} ({\bm x},Z,V_{z},t) \delta (Z-z)=
(1+\alpha_{P}^{-2} b_{P}^{-1}) \Phi_{+} ({\bm x},Z,V_{z},t)
\delta(Z-z).
\end{equation}
Here $b_{P}^{-1}$ is an operator acting on the velocities ${\bm v}$
and $V_{z}$ to its right, replacing them by the precollisional
values given by Eqs. (\ref{3.7})-(\ref{3.9}).

Equation (\ref{3.21}) shows that $\Phi_{0}$ is fully determined at
$z=Z$ if $\Phi_{+}$ is known at the same position. Consequently,
upon introducing simplifications or approximations on the value of
$\Phi_{0}$ at the piston, i.e. on $\Phi_{0} \delta(Z-z)$, they must
refer only to either $\Phi_{+} \delta(Z-z)$ or $\Phi_{-}
\delta(Z-z)$. Otherwise, the exact relationship given by Eq.
(\ref{3.21}) may be violated. A natural question in this context is
how relevant is that relation in practice. In other words, is any
fundamental physical property possibly lost if Eq. (\ref{3.21}) is
not satisfied in a given approximate description? The answer to
this question is affirmative as it will be seen in the next section.

In the limit of a very dilute gas, a simple approximation, similar
to the one leading to the Boltzmann equation \cite{Lu99}, is to
assume that
\begin{equation}
\label{3.22} \Phi_{+} ({\bm x},Z,V_{z},t) \delta(Z-z) = f({\bm x},t)
F(Z,V_{z},t) \Theta (g_{z}) \delta (Z-z),
\end{equation}
therefore neglecting all the  correlations between the piston and
the particles colliding with it {\em before} the collision. Using
the above approximation into Eq. (\ref{3.21}), it is found that
\begin{eqnarray}
\label{3.23} \Phi_{0} ({\bm x},Z,V_{z},t) \delta(Z-z) & = & f({\bm
x},t) F(Z,V_{z},t) \delta(Z-z) \nonumber \\
&& + \Theta (-g_{z}) (\alpha_{P}^{-2} b_{P}^{-1} -1 )  f({\bm x},t)
F(Z,V_{z},t) \delta(Z-z).
\end{eqnarray}
Upon deriving this equation use has been made of the relations
\begin{equation}
\label{3.24} b_{P}^{-1} \Theta (g_{z}) = \Theta (g_{z}^{*})
b_{P}^{-1} = \Theta (-g_{z}) b_{P}^{-1},
\end{equation}
following from the definition of $b_{P}^{-1}$ and the relationship
between $g_{z}$ and $g_{z}^{*}$. The physical meaning of Eq.
(\ref{3.23}) is evident: correlations between the piston and the
particles in its neighborhood are created by the collisions.

In this section, only the kinetic boundary condition for the piston
at the top of the system has been considered. The analysis of the
vibrating wall located at the bottom is much simpler, and it will be discussed later on.

\section{Hydrodynamic boundary conditions}
\label{s4} In the following, attention will be restricted to the
macroscopic steady state described in Secs.\ \ref{s2} and \ref{s3}.
Then, the height of the system $L$ there corresponds to the average
position of the piston, $\overline{Z}$, in the kinetic theory
description, while the average velocity of the piston is required to
vanish,
\begin{eqnarray}
\label{4.1} \overline{V}_{z} & \equiv & \frac{1}{N} \int d{\bm x}
\int_{0}^{\infty} dZ \int_{-\infty}^{\infty} dV_{z}\, V_{z} \Phi_{st}
({\bm
x},Z,V_{z}) \nonumber \\
&=& \int_{0}^{\infty} dZ \int_{-\infty}^{\infty}dV_{z}\, V_{z}
F_{st} (Z,V_{z}) =0,
\end{eqnarray}
where the indexes {\em st} are used to refer to properties of the
system in the steady state. Although the above property can be
accomplished in other ways, here the simplifying assumption that
$\Phi_{st}({\bm x},Z,V_{z})$ and, therefore, $F_{st}(Z,V_{z})$ are
even functions of $V_{z}$ is made. This is consistent with the
results from molecular dynamics simulations to be reported later on.
Also, it must be kept in mind that for the state we are considering,
$\Phi_{st}$ depends on the position of the particles only through
the coordinate $z$, although it is not made explicit in the
notation.

It is interesting and illuminating to compute the force ${\bm F}$
that the granular gas makes on the piston in this state. Because of
symmetry reasons, the force only has $z$-component, that is computed
in Appendix \ref{ap2} with the result
\begin{equation}
\label{4.2} F_{z}= W n_{L} T_{L,z},
\end{equation}
where $n_{L}$ is the number density of the granular gas next to the
piston, i.e. $n_{L}= n(z \rightarrow L)$, and $T_{L,z}$  is a
temperature parameter of the gas in the same region defined from the
$z$-component of the velocity or, equivalently, proportional to the
$zz$ component of the pressure tensor,
\begin{equation}
\label{4.3} T_{L,z} \equiv \frac{1}{n_{L}} \int dz \int d{\bm v}
\int _{0}^{\infty} dZ \int_{-\infty}^{\infty} dV_{z}\, m v_{z}^{2}
\Phi_{0,st}({\bm x}, Z,V_{z}) \delta(Z-z).
\end{equation}
In the Navier-Stokes approximation used in Sec. \ref{s1}, $T_{L,z}$
coincides with the temperature of the gas at the piston, $T_{L}$,
and Eq. (\ref{4.2}) is the expected result, since it agrees with the
hydrodynamic interpretation of the pressure tensor. In its
derivation, a crucial role is played by the kinetic boundary
condition given in Eq.\ (\ref{3.20}), as it is shown in Appendix
\ref{ap2}. Still more, the specific form of $\Phi_{+,st}$ is not
relevant, as long as $\Phi_{-,st}$ be consistently derived from it.
Otherwise, the pressure of the dilute gas defined as the product of
the local density times the local temperature, would not agree with
the scalar defined from the hydrodynamic pressure tensor,
characterizing the internal forces in the fluid.

In order to develop a consistent theory, it is then convenient to
express the hydrodynamic fields of the granular gas in the vicinity
of the piston in terms of only $\Phi_{+,st}$, instead of the
complete distribution $\Phi_{0,st}$. The expression of the number
density $n_{L}$ introduced above in terms of $\Phi_{0,st}$ reads
\begin{equation}
\label{4.4} n_{L} \equiv \int dz \int d{\bm v} \int_{0}^{\infty} dZ
\int_{-\infty}^{\infty} dV_{z}\, \Phi_{0,st}({\bm x},Z,V_{z})
\delta(Z-z).
\end{equation}
By means of the decomposition formulated in Eq.\ (\ref{3.16}) and
using the boundary condition (\ref{3.20}), the above expression can
be put in the form
\begin{equation}
\label{4.5} n_{L}= \frac{1+\alpha_{P}}{\alpha_{P}}\, n_{L}^{(+)},
\end{equation}
with
\begin{equation}
\label{4.6} n_{L}^{(+)} \equiv \int dz \int d{\bm v}
\int_{0}^{\infty} dZ \int_{-\infty}^{\infty} dV_{z}\,
\Phi_{+,st}({\bm x},Z,V_{z}) \delta(Z-z).
\end{equation}
Also, it is easily verified that the local velocity flow vanishes at
the piston in the steady state, as it should. Finally, consider the
temperature of the gas next to the piston, $T_{L}$, given by
\begin{equation}
\label{4.7} \frac{d}{2} n_{L} T_{L} \equiv \int dz \int d{\bm v}
\int_{0}^{\infty} dZ \int_{-\infty}^{\infty} dV_{z}\,
\frac{mv^{2}}{2}\, \Phi_{0,st}({\bm x},Z,V_{z}) \delta(Z-z).
\end{equation}
It is convenient to distinguish between the perpendicular and $z$
contributions to  this temperature,
\begin{equation}
\label{4.8} T_{L}=\frac{1}{d}\, T_{L,z} + \frac{d-1}{d}\,
T_{L,\perp},
\end{equation}
where $T_{L,z}$ is defined in Eq.\ (\ref{4.3}) and
\begin{equation}
\label{4.9} T_{L, \perp} \equiv \frac{1}{(d-1)n_{L}} \int dz \int
d{\bm v} \int _{0}^{\infty} dZ \int_{-\infty}^{\infty} dV_{z}\, m
v_{\perp}^{2} \Phi_{0,st}({\bm x}, Z,V_{z}) \delta(Z-z).
\end{equation}
The boundary condition (\ref{3.20}) leads directly to
\begin{equation}
\label{4.10} T_{L,\perp}=T_{L,\perp}^{(+)},
\end{equation}
with
\begin{equation}
\label{4.11} T_{L,\perp}^{(+)} \equiv \frac{1}{(d-1)n_{L}^{(+)}}
\int dz \int d{\bm v} \int_{0}^{\infty} dZ \int_{-\infty}^{\infty}
dV_{z}\, m v_{\perp}^{2} \Phi_{+,st}({\bm x},Z,V_{z}) \delta(Z-z).
\end{equation}
A more involved calculation gives
\begin{equation}
\label{4.12} T_{L,z} = \frac{\alpha_{P} (M-m)}{M- \alpha_{P}m}\,
T_{L,z}^{(+)}-H_{z}^{(+)},
\end{equation}
where
\begin{equation}
\label{4.13} T_{L,z}^{(+)} \equiv \frac{1}{n_{L}^{(+)}} \int dz \int
d{\bm v} \int_{0}^{\infty} dZ \int_{-\infty}^{\infty} dV_{z}\, m
v_{z}^{2} \Phi_{+,st}({\bm x},Z,V_{z}) \delta(Z-z),
\end{equation}
\begin{equation}
\label{4.14} H_{z}^{(+)} \equiv \frac{2
\alpha_{P}Mm}{(M-\alpha_{P}m)n_{L}^{(+)}} \int dz \int d{\bm v}
\int_{0}^{\infty} dZ \int_{-\infty}^{\infty} dV_{z}\, v_{z}V_{z}
\Phi_{+,st}({\bm x},Z,V_{z}) \delta(Z-z).
\end{equation}
At this point, a simplifying hypothesis on the pre-collisional
two-body distribution at contact is made. The associated marginal
velocity distribution is approximated by a product of Gaussian
distributions, namely it is assumed that
\begin{equation}
\label{4.15} \int dz \int_{0}^{\infty} dZ\, \Phi_{+,st}({\bm
x},Z,V_{z}) \delta(Z-z) = f_{st}^{(+)} ({\bm v}) P_{st}(V_{Z})
\Theta (g_{z}),
\end{equation}
where
\begin{equation}
\label{4.16} f_{st}^{(+)}({\bm v})= 2 n_{L}^{(+)} \varphi_{MB}({\bm
v}_{\perp}) \left( \frac{m}{2 \pi T_{L,z}^{(+)}} \right)^{1/2}
e^{-\frac{mv_{z}^{2}}{2T_{L,z}^{(+)}}},
\end{equation}
\begin{equation}
\label{4.17} \varphi_{MB}({\bm v}_{\perp})= \left( \frac{m}{2 \pi
T_{L,\perp}} \right)^{(d-1)/2} e^{- \frac{m
v_{\perp}^{2}}{2T_{L,\perp}}},
\end{equation}
\begin{equation}
\label{4.18} P_{st}(V_{z}) = \left( \frac{M}{2 \pi T_{P}}
\right)^{1/2} e^{- \frac{MV_{z}^{2}}{2 T_{P}}}.
\end{equation}
Although in the same spirit, Eq. (\ref{4.15}) is a somewhat stronger
assumption than the particularization of Eq.\ (\ref{3.22}) for the
steady state under consideration. The presence of $n_{L}^{(+)}$ and
$T_{L,z}^{(+)}$ in Eq.\ (\ref{4.16}) and of $T_{L,\perp}$ in Eq.\
(\ref{4.17}) is required by consistency with the previous results in
this section. Moreover, it will be assumed in the following that
$T_{L,\perp}=T_{L}$ for the sake of simplicity. Because of Eq.\
(\ref{4.8}) this implies that also $T_{L,z}=T_{L}$ and, therefore,
all the diagonal components of the pressure tensor of the gas in the
vicinity of the piston are the same, consistently with the
Navier-Stokes approximation.

Using Eqs.\ (\ref{4.15})-(\ref{4.18}) it is obtained that
$H_{Z}^{(+)}=0$ and, therefore, Eq. (\ref{4.12}) reduces to
\begin{equation}
\label{4.19} T_{L}=T_{L,z} = \frac{\alpha_{P}(M-m)}{M-\alpha_{P}m}\,
T_{L,z}^{(+)}.
\end{equation}
Equations (\ref{4.5}) and (\ref{4.19}) relate the properties of the
flux of grains reaching the wall with the local hydrodynamic fields
of the granular gas next to the wall. They will be employed now to
derive an expression for the heat flux at the piston, $Q_{L}$. Using
standard kinetic theory arguments, this quantity can be computed
from the variation of kinetic energy of the grains colliding with
it, namely it is given by the energy of the flux leaving the piston
minus that of the flux reaching it. Then, it can be written as
\begin{equation}
\label{4.20} Q_{L} = \int dz \int d{\bm v} \int_{0}^{\infty} dZ
\int_{-\infty}^{\infty} dV_{z}\ \frac{m}{2} (v_{z}^{2}-v_{z}^{\prime
2}) g_{z}  \Phi_{+,st} ({\bm x},Z,V_{z}) \delta (Z-z).
\end{equation}
This expression is easily evaluated in the Gaussian approximation
given by Eqs.\ (\ref{4.15})-(\ref{4.18}), with the result
\begin{equation}
\label{4.21} Q_{L} = 2 \left( \frac{2}{\pi m}\right)^{1/2}
\frac{M}{M+m} \left( \frac{1+\phi}{\alpha_{P}}\right)^{1/2} \left(
\frac{M-\alpha_{P} m}{M-m} \right)^{3/2} \left[ 1-
\frac{(1+\alpha_{P})(1+\phi) M}{2(M+m)}\right] n_{L} T_{L}^{3/2},
\end{equation}
where
\begin{equation}
\label{4.22} \phi \equiv \frac{m T_{P}}{MT_{L,z}^{(+)}} =
\frac{\alpha_{P}(M-m)m T_{P}}{(M-\alpha_{P}m)MT_{L}}\, .
\end{equation}
The calculation of the power injected into the system through the
vibrating wall at the bottom is much more direct and it is outlined
in Appendix \ref{ap3}. There, it is shown that the heat  flux at
this wall is given by
\begin{equation}
\label{4.23} Q_{0}= v_{W} p_{0},
\end{equation}
with $p_{0} \equiv p(z=0) = p (\xi = \xi_{M})$ is the pressure of
the granular gas in the region just above the vibrating wall.
Equation $(\ref{4.23})$ was proposed in refs. \cite{McyB97} and
\cite{McyL98}, and used many times in the literature since then. To
get the value of $p_{0}$, first note that Eq.\ (\ref {2.13a}) gives
\begin{equation}
\label{4.24} p_{0}= p_{L}+mg_{0} N_{z}.
\end{equation}
The value of $p_{L}$  can be obtained by means of Eq. (\ref{4.2}),
after substituting $T_{L,z}$ by $T_{L}$ accordingly with Eq.
(\ref{4.19}). Mechanical equilibrium of the piston in the steady
state requires that $F_{z}=Mg_{0}$ and, therefore,
\begin{equation}
\label{4.25} p_{L} \equiv n_{L}T_{L} = \frac{M g_{0}}{W}\, .
\end{equation}
Once $p_{L}$ is known, the values of $\xi_{M}$ and $\xi_{m}$ defined
in Eqs. (\ref{2.12}) and (\ref{2.13}), respectively, can be
determined. The next step is to identify the constants $A$ and $B$
introduced in Sec. \ref{s2}, and present in the expressions of the
density and temperature profiles, Eqs. (\ref{2.16}) and
(\ref{2.18}). To do so, two boundary conditions are needed. They are
provided by requiring the values of $Q_{L}$ and $Q_{0}$ given above
to agree with the hydrodynamic expression of the heat flux $q_{z}$
given in Eq.\ (\ref{2.2a}), particularized for each of the wall
boundaries, i.e.,
\begin{equation}
\label{4.26} Q_{L}=\lim_{z\rightarrow Z} q_{z}(z),
\end{equation}
\begin{equation}
\label{4.27} Q_{0} = \lim_{z \rightarrow 0} q_{z}(z),
\end{equation}
with
\begin{equation}
\label{4.28} q_{z}(z) = -\kappa_{0} \left[ (\kappa^{*}- \mu^{*})
\frac{\partial T}{\partial z} - \mu^{*} m g_{0} \right].
\end{equation}
By using Eqs. (\ref{2.9}), (\ref{2.16}), and (\ref{2.17}), as well
as properties of the modified Bessel functions \cite{AyS65}, the
above expression becomes
\begin{equation}
\label{4.29} q_{z}(\xi)= 2 m g_{0} \frac{\kappa_{0}(T)}{T^{1/2}}\, (
\kappa^{*}-\mu^{*}) \xi^{1-\nu} \left[ A I_{\nu-1}(\xi)-B K
_{\nu-1}(\xi) \right].
\end{equation}
Making in this equation $\xi=\xi_{m}$ and equating the result to the
expression of $Q_{L}$ in Eq. (\ref{4.21}) it is obtained:
\begin{equation}
\label{4.30} \frac{B}{A} = \frac{ I_{\nu-1} (\xi_{m})-e(\alpha,
\alpha_{P}) I_{\nu}(\xi_{m}) }{K_{\nu-1} (\xi_{m})+e(\alpha,
\alpha_{P}) K_{\nu}(\xi_{m})},
\end{equation}
where
\begin{eqnarray}
\label{4.31} e(\alpha,\alpha_{P}) & =  & \frac{ 16 \sqrt{2} (d-1)
\pi^{(d-2)/2}}{d(d+2)^{2} \Gamma (d/2)}\, \frac{M}{M+m} \left(
\frac{1+\phi}{\alpha_{P}}\right)^{1/2}
\nonumber \\
&& \times \left( \frac{M-\alpha_{P} m}{M-m} \right)^{3/2} \left[ 1-
\frac{(1+\alpha_{P})(1+\phi) M}{2(M+m)}\right]
\frac{1}{\sqrt{a(\alpha)} (\kappa^{*}- \mu^{*})}\, .
\end{eqnarray}

Next, Eq.\ (\ref{4.29}) is particularized for $\xi = \xi_{M}$ and
afterwards put equal to the right hand side of Eq.\ (\ref{4.23}).
This leads to
\begin{equation}
\label{4.32} AI_{\mu-1}(\xi_{M}) - B K_{\mu -1} (\xi_{M}) =
\xi_{M}^{\nu} \frac{8(d-1) \pi^{(d-1)/2}}{d(d+2)^{2} \Gamma (d/2)}\,
\frac{m^{1/2} v_{W}}{\sqrt{a(\alpha)} (\kappa^{*}-\mu^{*})}\, .
\end{equation}

An alternative to one of the two boundary conditions used above
would be to require that the total power dissipated in the system
due to the inelasticity of collisions be the same as the net heat
flux injected in the system through the boundaries. This leads to a
relation between $A$ and $B$ that is a combination of Eqs.
(\ref{4.30}) and (\ref{4.32}), consistently with the fact that the
condition follows directly from the hydrodynamic equations. This can
be realized by noting that the energy balance equation in the steady
state reads (see Eqs. (\ref{2.2}) and (\ref{2.2a}))
\begin{equation}
\label{4.33} \frac{\partial q_{z}}{\partial z}= -\frac{nd}{2}\ T
\zeta^{(0)},
\end{equation}
whose integration between $z=0$ and $z=L$ gives
\begin{equation}
\label{4.34} q_{z,L}-q_{z,0}= - \frac{d}{2} \int_{0}^{L} dz\, p(z)
\zeta (z).
\end{equation}
The right hand side of this equation is clearly identified as the
energy dissipated in the system per unit of time and section due to
the inelasticity of collisions.

\begin{figure}
\includegraphics[scale=0.7,angle=0]{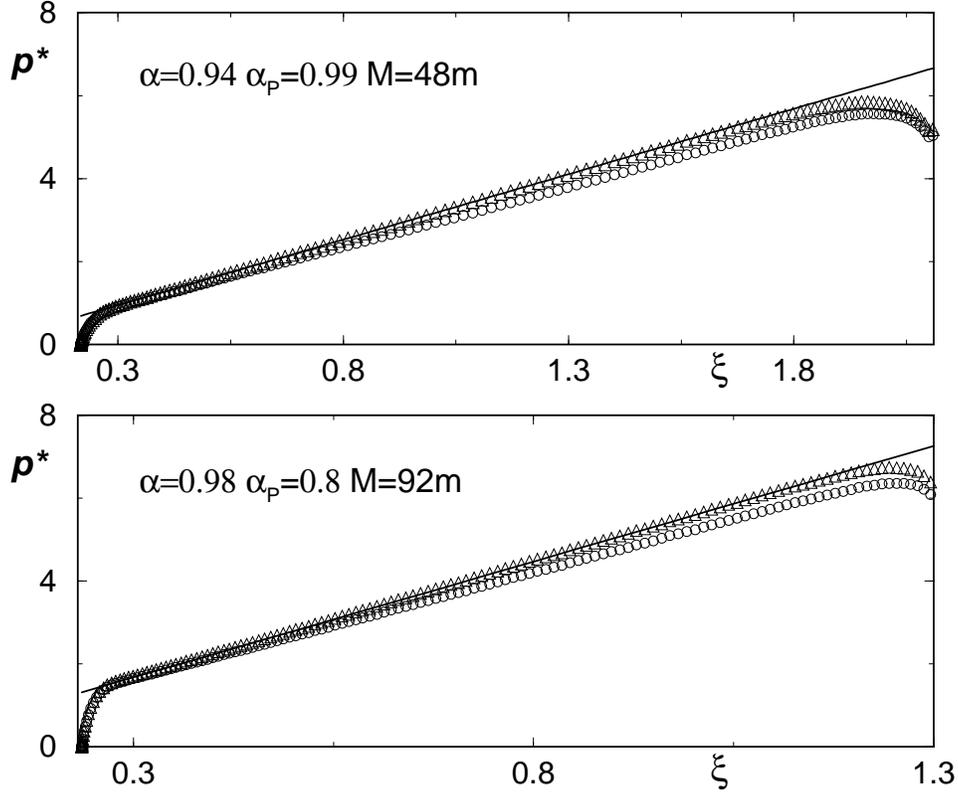}
\caption{Dimensionless pressure ($p^{*}$) profiles for two steady
states of the system of inelastic hard disks depicted in Fig.\
\protect{\ref{fig1}}. The solid lines are the theoretical
predictions given in the main text, while the symbols are MD
simulation results. The empty circles have been computed using
$p^{*}=n^{*}T^{*}$, while the triangles have been calculated by
means of Eq. (\protect{\ref{5.2}}). The values of the parameters of
the system are $M=48m$, $\alpha=0.94$, $\alpha_{P} = 0.99$,  $v_{W}
= 7 \sqrt{g_{0} \sigma}$ for the top figure, and $M=92m$,
$\alpha=0.98$, $\alpha_{P} = 0.8$,  $v_{W} = 3 \sqrt{g_{0} \sigma}$,
for the bottom one }\label{fig2}
\end{figure}

\begin{figure}
\includegraphics[scale=0.7,angle=0]{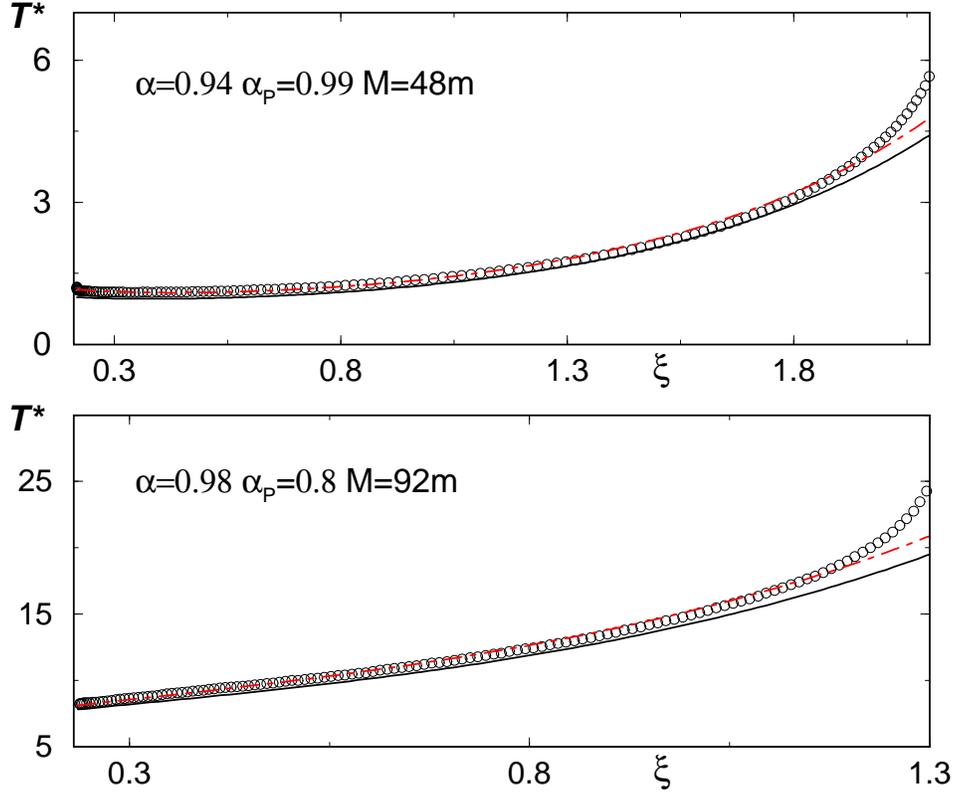}
\caption{(Color online) Dimensionless temperature profile $T^{*}$ for the same two
steady states as in Fig. \protect{\ref{fig2}}. The solid line is the
theoretical prediction given by Eq.\ (\protect{\ref{2.16}}) with the
values of $A$ and $B$ determined in Sec. \protect{\ref{s4}}. The
empty circles are MD results and the dashed line (red online) is the
best fit to Eq.\ (\protect{\ref{2.16}}) by varying $A$ and $B$.}
\label{fig3}
\end{figure}

To close the identification of the hydrodynamic profiles in the
system, a prescription to determine the parameter $\phi$ defined in
Eq. ({\ref{4.22}) is needed. This in turn calls for an expression
for the temperature of the piston $T_{P}$. It is well known that a
peculiar feature of granular gases is the violation of energy
equipartition \cite{GyD99,DHGyD02,WyP02,FyM02}, that in the present
case manifests itself by the difference between $T_{P}$ and $T_{L}$.
This has been discussed with some detail in \cite{ByR08}. For values
of the restitution coefficient $\alpha$ close to unity, a good
approximation to the value of $\phi$ is obtained by taking
$T_{P}/T_{L}$ equal to unity.

A relevant quantity characterizing the macroscopic state of the
system is the average height of the piston, $L= \overline{Z}$. The
theoretical prediction for it is obtained from Eq.\ (\ref{2.19})
once the values of $\xi_{M}$ and $\xi_{m}$ have been obtained and
the density profile is known,
\begin{equation}
\label{4.35} L = \frac{1}{\sqrt{a(\alpha)} \sigma^{d-1}}
\int_{\xi_{m}}^{\xi_{M}} \frac{d \xi}{n(\xi)}\, .
\end{equation}

\section{Molecular Dynamics simulations}
\label{s5} To test the theoretical predictions discussed in the
previous sections, molecular dynamics (MD) simulations of a
two-dimensional ($d=2$) system of inelastic hard disks have been
performed. To avoid undesired boundary effects, periodic boundary
conditions with periodicity $W$ were used in the direction
perpendicular to the $z$ axis. In all the simulations to be reported
in the following, the grains were initially located in a square
lattice with a Gaussian velocity distribution. The initial position
of the piston was slightly above the highest layer of grains. Then,
the system was evolved in time accordingly with the mechanical rules
governing the dynamics of the particles and the piston, and it was
observed that, for values of the restitution coefficient $\alpha$
between $0.9$ and $1$, a steady state with only gradients in the
vertical direction and no flow field  was reached. Actually, in
order for the system to reach this state, its width $W$ has to be
not too large, since otherwise transversal inhomogeneities of the
type discussed in refs. \cite{BRMyG02} and \cite{LMyS02} develop
into the system. The results presented below have been time averaged
once the system is in the steady state, and also over several
trajectories. More precisely, once the stationary state was reached,
its trajectory was followed for $4000$ collisions per particle, and
$10$ different trajectories were used in each case.

To carry out a systematic analysis of the theoretical predictions,
it is necessary to reduce the number of parameters of the system by
fixing some of them. In the present study, the values $N=420$ and
$W=70 \sigma$ ($N_{z}=6 \sigma^{-1}$) have been used in all the
simulations. The reason is that the dependence on $N_{z}$ of the
theoretical results follows trivially once the hydrodynamic
description is assumed to hold, and analysis of the simulation
results indicates that the above values are appropriate for the
purposes here, in the sense that a hydrodynamic behavior is
observed. Moreover, only results for values of $\alpha$ in the
interval $0.9 \leq \alpha < 1$ will be reported, the reason being
that for smaller values of the restitution coefficient of the gas,
qualitative and quantitative strong deviations from the theoretical
predictions were found. This is not surprising, since the
Navier-Stokes approximation is expected to fail beyond the weak
dissipation limit, as a consequence of the coupling between
gradients and inelasticity, as pointed out in Sec. \ref{s2}.

\subsection{Hydrodynamic profiles}
In Figs. \ref{fig2} and \ref{fig3}, the dimensionless pressure
($p^{*}$) and temperature ($T^{*}$) profiles, respectively, are
plotted as a function of the length scale $\xi$ defined in Eq.\
(\ref{2.9}), for two different combinations of the parameters $M/m$,
$\alpha$, and $\alpha_{P}$, as indicated in the figures themselves.
Remind that $\xi$ is a {\em decreasing} function of $z$. The
dimensionless fields are defined by
\begin{equation}
\label{5.1} T^{*} \equiv  \frac{T}{m v_{W}^{2}}, \quad p^{*} \equiv
\frac{p m \sigma}{g_{0}}\, .
\end{equation}
It is easily seen that in these units, the theoretical predictions
for the hydrodynamic fields become independent of the velocity
$v_{W}$  of the vibrating wall and also of the acceleration $g_{0}$.
Of course, this applies as long as the system is fluidized and its
density low everywhere.

Consider first the pressure field. The theoretical prediction is
given by Eq.\ (\ref{2.13a}) and does not involve the parameters $A$
and $B$ appearing in the temperature and density profiles. It is
represented by the solid lines in Fig.\ \ref{fig2}. In the
simulations, the $\xi$ coordinate has been measured directly by
using the discrete version of Eq.\ (\ref{2.9}). On the other hand,
the pressure at each value of $z$ has been obtained from the data
for the density and the temperature at the same value. In the low
density limit considered in the theory, it is $p^{*}=n^{*}T^{*}$,
where $n^{*} \equiv n \sigma v_{W}^{2} /g_{0}$. The results for the
pressure represented by empty circles in the figures have been
computed in this way. A good agreement is observed between theory
and simulation in the bulk of the system, i. e. outside the boundary
layers next to the upper and lower walls. The results can be
considered as satisfactory, especially taking into account that
there are no adjustable parameters. Nevertheless, it is true that a
systematic although small deviation is observed. Equation
(\ref{2.13a}) is a quite general result, which only requires for its
derivation the restriction to the Navier-Stokes order approximation,
without any particular expression of the transport coefficients or
of the (local) equation of state. Thus it seems sensible to check
whether the origin of the observed discrepancy lies in the way in
which the pressure is computed from the simulation data, namely by
using the ideal gas equation of state. For this reason, the local
pressure has also been calculated from the MD data by using the
equation of state proposed by Grossman et {\em al}. \cite{GZyB97}}
for hard disks at finite density,
\begin{equation}
\label{5.2} p= n T \frac{n_{c}+n}{n_{c}-n}\, ,
\end{equation}
where $n_{c} = 2 \sigma^{-2}/\sqrt{3}$ is the maximum packing number
density. The pressure values obtained by this procedure  are
represented by empty triangles in Fig. \ref{fig2}. Now, a much
better agreement is obtained. It is concluded that density
corrections are identifiable in the pressure, in spite of the
density being quite small. Actually, the same values of the pressure
are obtained if instead of Eq. (\ref{5.2}), the second virial
approximation for the equation of state of a gas of hard disks is
used.

The scaled temperature profiles shown in Fig. \ref{fig3} also
exhibit a good agreement between the theoretical predictions and the
simulation results. Moreover, the best fits obtained varying the
parameters $A$ and $B$ in Eq. (\ref{2.16}) are also plotted.
Introducing density corrections to the expression for the
temperature as discussed above for the pressure, is far from
trivial. The arguments in Sec. \ref{s2} leading to a closed
separated equation for the temperature do not apply anymore and the
resulting differential equations do not seem to have an elementary
solution. In any case, the above results clearly confirm the
accuracy of both the hydrodynamic equations and the used boundary
conditions. Notice that the boundary layer next to the movable
piston is much narrower for the temperature than for the pressure.

Similar degree of agreement has been found for all the studied
combinations of parameters in the intervals $20 \leq M/m \leq 120$,
$0.9 \leq \alpha < 1$, and $0.6 \leq \alpha_{P} \leq 1$. In
particular, the inelasticity of the collisions between the movable
piston and the grains, measured by the coefficient $\alpha_{P}$,
seems to affect very weakly the accuracy of the theory.

\begin{figure}
\includegraphics[scale=0.7,angle=0]{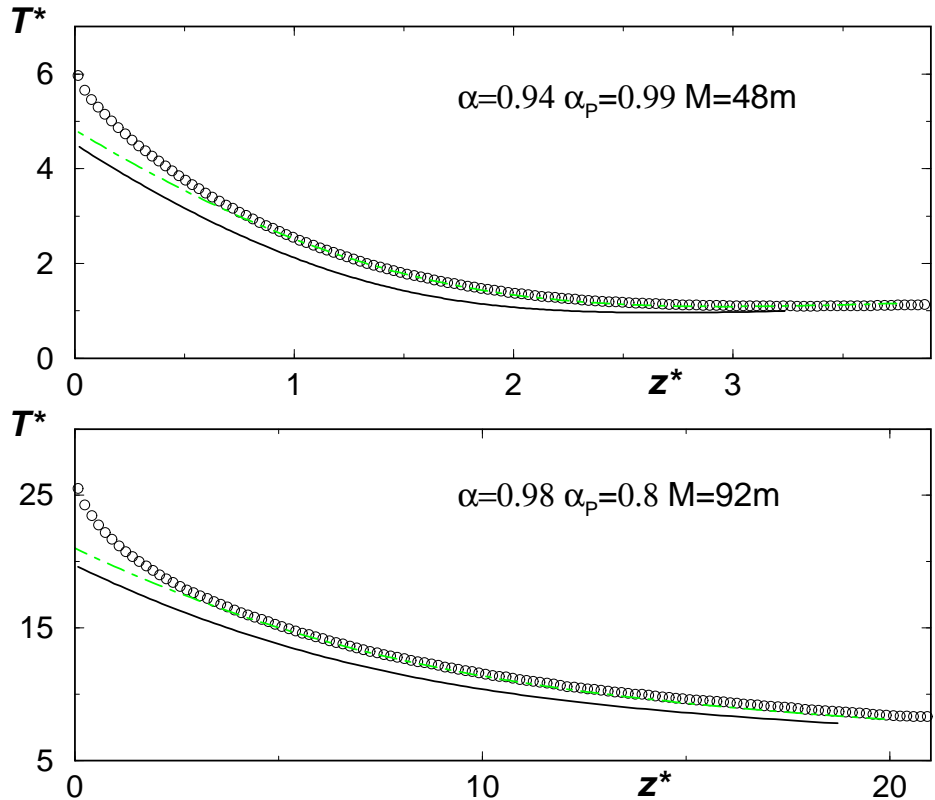}
\caption{(Color online) Temperature profiles for the same systems as in Fig.
\protect{\ref{fig2}}, but now as a function of the reduced original
coordinate $z^{*}$ defined in the text. The solid line is again the
theoretical prediction with no adjustable parameters, and the
symbols MD simulation results. The dotted-dashed lines (green
online) have been obtained by fitting the two parameters $A$ and $B$
appearing in the expression of the temperature field using,
moreover, the equation of state (\protect{\ref{5.2}}), as discussed
in the main text.} \label{fig4}
\end{figure}

\begin{figure}
\includegraphics[scale=0.7,angle=0]{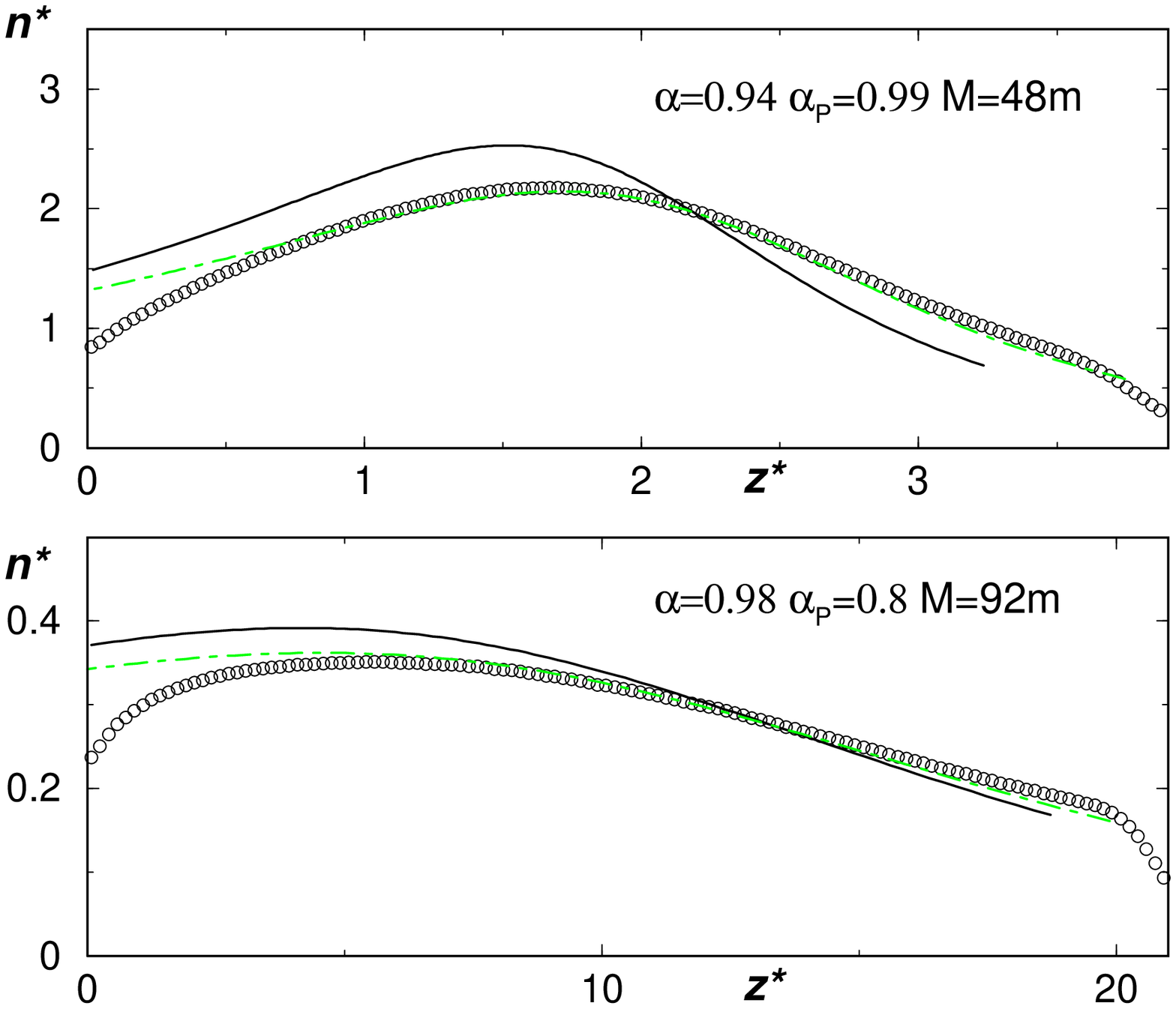}
\caption{(Color online) Density profiles for the same systems as in Fig.
\protect{\ref{fig2}} as a function of the reduced  coordinate $z^{*}
$ defined in the text. The solid line is the theoretical prediction
with no adjustable parameters and the symbols MD simulation results.
The dotted-dashed lines (green online) have been obtained by fitting
the two parameters $A$ and $B$ appearing in the expression of the
temperature field using, moreover, the equation of state
(\protect{\ref{5.2}}), as discussed in the main text.} \label{fig5}
\end{figure}

It is interesting to plot also the hydrodynamic profiles in the
original scale $z$ or, equivalently as a function of $z^{*}=z
g_{0}/v_{W}^{2}$. In Fig. \ref{fig4}, the temperature profiles in
the $z^{*}$-scale are shown for the same two steady states as in the
previous figures. Also included are the profiles obtained with the
best fitting parameters $A$ and $B$. It is observed that the
agreement between theory and simulation is now worse than in the
previous figures, where the spatial scale $\xi$ was employed. This
is not at all surprising, since in the transformation of the
theoretical prediction from the scale $\xi$ to the scale $z$, Eq.
(\ref{2.9}), the density profile is involved, and the ideal gas
equation of state has been used. Consequently, the density effects
discussed above manifest themselves in the transformation. Moreover,
the scale $z$ is defined from $ \xi$ in a cumulative way, as an
integral over the density profile, so that the discrepancies
increase as $z$ decreases. This effect is also clearly identified in
Fig. \ref{fig5}, where the profiles for the reduced density $n^{*}$
are given. It is also seen that the simulation results extend to
larger values of $z$ than the theoretical predictions for the
profiles. It is worth to emphasize that the maximum value of the
density, $n_{\max}$, remains quite  low in both systems. For the system with $\alpha=0.94$,
it is $n_{\max} \simeq 0.045 \sigma^{-2} \simeq 0.039 n_{c}$, while for the one
with  $\alpha = 0.98$, $n_{\max} \simeq 0.04 \sigma^{-2} \simeq 0.035 n_{c}$.

The previous analysis of the profiles in the $\xi$ variable suggests
that the analytical expression of the reduced temperature in that
variable is rather robust, in the sense of being very little
affected by the finite density effects that, on the other hand,
produce identifiable modifications on the pressure and density
profiles. To check this idea, which can be useful to describe real
experiments, the simulation profiles in Figs. \ref{fig4} and
\ref{fig5} have been also fitted as follows. The pressure is given
by Eq.\ (\ref{2.13a}), and for the temperature, $T(\xi)$, Eq.\
(\ref{2.16}) is used, with $A$ and $B$ being adjustable parameters.
Nevertheless, the equation of state is not that of an ideal gas, but
Eq.\ (\ref{5.2}). This equation of state is used both to compute the
pressure from the simulation data, as discussed above, and also to
transform from the $\xi$ coordinate to the $z$ one by means of Eq.
(\ref{2.9}). The profiles obtained in this way are also included in
Figs. \ref{fig4} and \ref{fig5}, and reproduce fairly well the
hydrodynamic profiles in the bulk on the system.

\subsection{Global properties and boundary conditions}

\begin{figure}
\includegraphics[scale=0.7,angle=0]{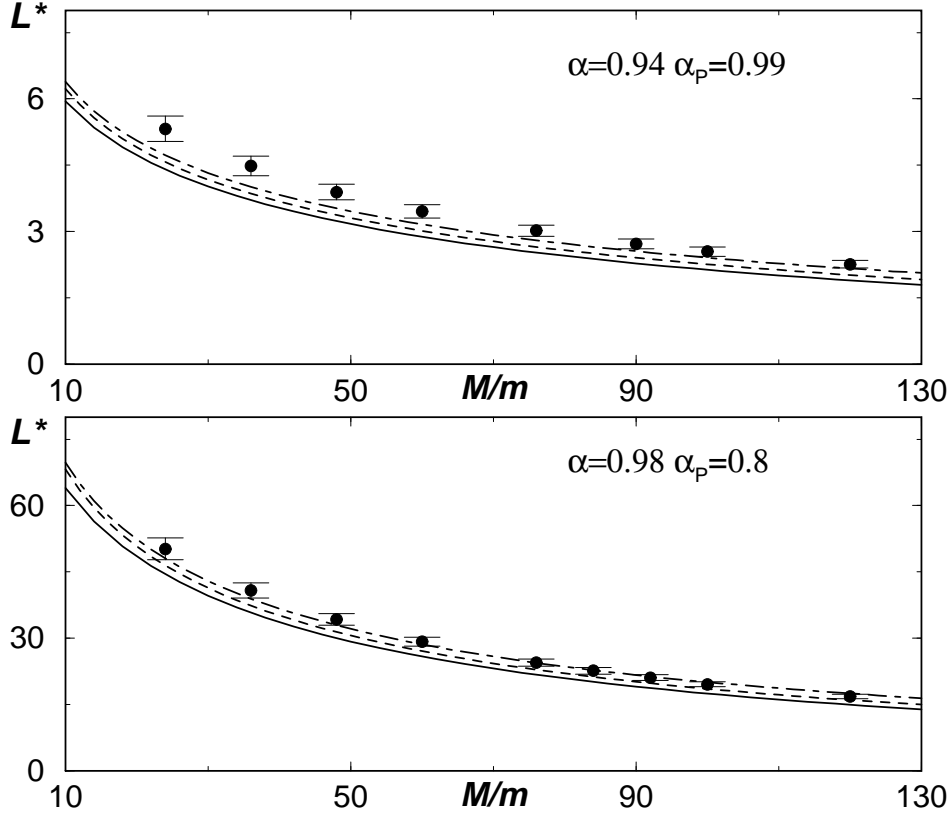}
\caption{ Average position of the piston $L^{*}$, measured in the
dimensionless units defined in the text, as a function of the the
mass ratio $M/m$, for two different pairs of the restitution
coefficients $\alpha$, $\alpha_{P}$, as indicated in the figures.
The symbols are MD simulation results and the solid lines the
theoretical prediction derived in Sec.\ \protect{\ref{s4}}. The
dashed lines and the dot-dashed lines have been obtained by using
the equation of state (\protect{\ref{5.2}}), and correspond to the
largest and smallest values of the velocity of the bottom wall
$v_{W}$ used in the simulation, respectively.} \label{fig6}
\end{figure}

The theoretical result for the average position of the piston $L$ is
given by Eq. (\ref{4.35}), with the density profile given by Eq.\
(\ref{2.18}) and the values of the constants $A$ and $B$ following
from Eqs.\ (\ref{4.30}) and (\ref{4.32}). In Fig. \ref{fig6}, the
above prediction is compared with MD simulation results. The value
of the dimensionless height $L^{*}=L g_{0} /v_{W}^{2}$ as a function
of $M/m$ is plotted for two systems, one with $\alpha=0.94$,
$\alpha_{P}= 0.99$, and the other one with $\alpha=0.99$,
$\alpha_{P}=0.8$. The error bars in the simulations data have been
obtained from the mean square fluctuations of the position of the
piston around its average value, once in the steady state. Although
there is a systematic underestimation of $L$, the agreement between
theory and simulation is good over a quite wide range of values of
the mass ratio. Prompted by the previous discussion of the
hydrodynamic profiles, it is tempting to investigate whether some of
the observed discrepancy is due to finite density corrections. Then,
instead of using Eq. (\ref{2.18}) for the density profile, the
equation of state (\ref{5.2}) was employed. In the latter, the
pressure was given by Eq. (\ref{2.13a}) and the temperature by Eq.\
(\ref{2.16}), with $A$ and $B$ still determined from Eqs.\
(\ref{4.30}) and (\ref{4.32}). The problem is that now the resulting
expression for $L^{*}$ depends on the velocity of the bottom
vibrating wall, $v_{W}$, and the simulation data in Fig.
(\ref{fig6}) have been obtained with different values of this
velocity. For this reason, two different results are reported in the
figure, corresponding to the two extreme values of $v_{W}$  used in
the simulations. Although the dependence of the results on $v_{W}$
is small, it is clearly appreciable on the scale of the figure. As
expected, the curve corresponding to the largest value of $v_{W}$ is
the closest to the ideal gas prediction, since the larger $v_{W}$
the more dilute the system. Although some discrepancy between theory
and simulation still persists, increasing as the value of $\alpha$
decreases, it seems fair to conclude the presence of finite density
effects in the simulation results.

Another relevant property characterizing globally the system is the
power dissipated per unit of section $W$, as a consequence of the
inelasticity of collisions. As discussed at the end of Sec.
\ref{s4}, this quantity is related with the heat flux at the
boundaries  by
\begin{equation}
\label{5.3} D=Q_{L}-Q_{0}.
\end{equation}
The fluxes $Q_{L}$ and $Q_{0}$ have been computed in Sec.\ \ref{s4}
by using approximate kinetic theory arguments and the expressions
derived for them are given by Eqs. (\ref{4.21}) and (\ref{4.23}),
respectively. These expressions are closed by means of Eqs.
(\ref{4.24}) and (\ref{4.25}) and making, as above, the additional
approximation $T_{P}/T_{L} \simeq 1$ in the expressions of $\phi$,
Eq.\ (\ref{4.22}).

\begin{figure}
\includegraphics[scale=0.7,angle=0]{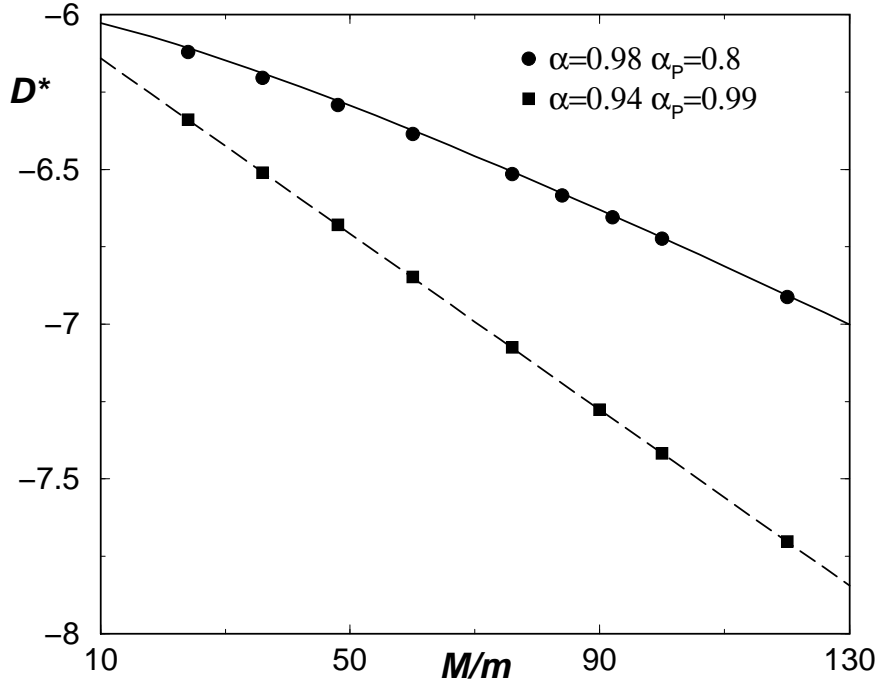}
\caption{Dimensionless dissipated power $D^{*}$ per unit of
transversal length in the steady state, as a function of the ratio
between the mass of the piston, $M$,  and the mass of the grains,
$m$. The symbols are MD simulation results, while the lines have
been obtained from the expressions of the fluxes at the boundaries
derived in this paper. The circles and solid line correspond to
systems with $\alpha=0.98$ and $\alpha_{P}=0.8$, and the squares and
the dashed line to $\alpha=0.94$ and $\alpha_{P}=0.99$.}
\label{fig7}
\end{figure}

In the MD simulations, $D$ is directly obtained from the loss of
kinetic energy in each collision. In Fig.\ \ref{fig7},
\begin{equation}
\label{5.4} D^{*} \equiv \frac{\sigma D }{m g_{0} v_{W}}
\end{equation}
is plotted as a function of $M/m$ for the same pairs of values
$\alpha, \alpha_{P}$ as in Fig.\ \ref{fig6}. The agreement between
theory and MD simulation is quite good. This provides strong
additional support for the analysis of the boundaries carried out in
Secs. \ref{s3} and \ref{s4} and, consequently, for the boundary
conditions employed in the hydrodynamic equations.

\section{Conclusions}
\label{s6} In this paper, the steady state of a fluidized granular
gas of smooth inelastic hard spheres in presence of gravity, and
with a movable lid on the top, has been investigated. It has been
shown that, for weak inelasticity of the grain-grain collisions, the
inelastic Navier-Stokes equations provide an accurate description of
the density, temperature, and pressure profiles. For stronger
inelasticity of the grains, a hydrodynamic description beyond the
Navier-Stokes approximation is needed due to the coupling between
gradients and inelasticity.

The issue of the required boundary conditions to solve the
hydrodynamic equations has also been addressed. Here, the system was
fluidized by means of a vibrating elastic wall at the bottom. For
simplicity, it was assumed to move in a sawtooth way and with high
frequency and small amplitude. The collisions of the particles with
the piston on the top were modeled as smooth and inelastic. The
above mechanical characterization of the walls was translated into
exact boundary conditions for the joint distribution functions of
each of the walls and the gas next to it. These kinetic boundary
conditions  are general, with restriction  neither in the gas
density nor in the inelasticity. They must be taken into account
when finding approximate hydrodynamic boundary conditions.
Otherwise, relevant exact properties, as the hydrodynamic meaning of
the pressure of the gas at the boundaries, can be violated. Here, an
approximation consistent with the kinetic boundary condition was
developed and shown to lead to quite accurate results. It must be
emphasized that the aim here has not been to analyze in detail the
boundary layers next to the walls, but rather to identify the
effective boundary conditions that must be satisfied by the
hydrodynamic fields extrapolated to the boundaries from the bulk of
the system.

Very recently, the so-called Knudsen temperature jump at a thermal
wall has been incorporated at the Navier-Stokes equations of a
weakly inelastic dilute gas of hard disks \cite{KMyS08}. As
mentioned in the Introduction, thermal walls have not been proven to
correspond to any well defined limit of vibrating walls. For the
specific case of the inelastic piston on top of the system
considered here, any analogy with a thermal wall seems very hard to
justify. Moreover, the approximation followed in \cite{KMyS08}
involves the determination of a constant prefactor by means of MD
simulations, while here all the constants are determined by the
theory.

The accuracy of the theoretical predictions has been  checked by
comparison with MD simulations. For some of the properties measured,
e.g. the pressure profile, finite density effects have been
identified and a practical way of accounting for them has been
proposed. Of course, these effects could have been avoided by using
the direct simulation Monte Carlo (DSMC) method \cite{Bi94,BRyC96},
instead of molecular dynamics.  Nevertheless,  the latter seems more
appropriate in the present context, in which the interest focuses
on the hydrodynamic description of a  given state and not on a
property of the Boltzmann equation. MD simulations provide results
closer to real experiments since they are not based on the validity
of any kinetic equation.

The interest has been on the description of the granular gas and its
interaction with the piston. Of course, the study of the properties
of the piston itself is also of great interest. For instance, its
position fluctuations give the volume fluctuations of the gas. Some
partial results have already been reported elsewhere \cite{ByR08},
but many points still deserve additional attention.

\section{Acknowledgements}

This research was supported by the Ministerio de Educaci\'{o}n y
Ciencia (Spain) through Grant No. FIS2008-01339 (partially financed
by FEDER funds).

\appendix

\section{Transport coefficients for a dilute granular gas}
\label{ap1} Here, the expressions of the dimensionless functions
introduced in Sec. \ref{s2} are given for the sake of completeness.
They read \cite{BDKyS98,ByC01}
\begin{equation}
\label{ap1.1} \kappa^{\ast}(\alpha)=
[\nu^{\ast}_{2}(\alpha)-\frac{2d}{d-1}\zeta^{\ast}(\alpha)]^{-1}
[1+c^{\ast}(\alpha)],
\end{equation}
\begin{equation}
\label{ap1.2} \mu^{\ast}(\alpha) =
2\zeta^{\ast}(\alpha)\left[\kappa^{\ast}(\alpha)+
\frac{(d-1)c^{\ast}(\alpha)}{2d\zeta^{\ast}(\alpha)} \right] \left[
\frac{2(d-1)}{d}\nu^{\ast}_{2}(\alpha)-
3\zeta^{\ast}(\alpha)\right]^{-1} \,  ,
\end{equation}
\begin{equation}
\label{ap1.3} \zeta^{\ast}(\alpha) =\frac{2+d}{4d}(1-\alpha^{2})
\left[1+\frac{3}{32} c^{*}(\alpha) \right]\, ,
\end{equation}
where
\begin{equation}
\nu_{2}^{\ast}(\alpha) \equiv \frac{1+\alpha}{d-1}\left[
\frac{d-1}{2}
+\frac{3(d+8)(1-\alpha)}{16}+\frac{4+5d-3(4-d)\alpha}{1024}
c^{\ast}(\alpha)\right],
\end{equation}
\begin{equation}
c^{\ast}(\alpha) \equiv
\frac{32(1-\alpha)(1-2\alpha^{2})}{9+24d+(8d-41)\alpha+
30\alpha^{2}(1-\alpha)}\, .
\end{equation}
These results have been obtained from the inelastic Boltzmann
equation for hard spheres and disks by means of a generalized
Chapman-Enskog algorithm, in the so-called first Sonine
approximation \cite{BDKyS98,ByC01}.

\section{Calculation of the force exerted by the gas on the piston in the steady state}
\label{ap2}

The force exerted by the granular gas on the piston equals the rate
of momentum transferred from the gas to the piston in the
collisions. When a particle collides with the piston, the amount of
momentum given by the former to the latter is given by Eq.\
(\ref{3.5}). Then, the average force on the piston $F_{z}$ is
\begin{eqnarray}
\label{ap2.1} F_{z} & = & \int d{\bm x} \int_{0}^{\infty} dZ
\int_{-\infty}^{\infty}dV_{z}\, \Theta (g_{z}) g_{z} \Delta P_{z}
\Phi_{0,st}({\bm x},Z,V_{z}) \delta(Z-z) \nonumber \\
&=& \frac{mM}{m+M}\, (1+\alpha_{P}) \int d{\bm x} \int_{0}^{\infty}
dZ \int_{-\infty}^{\infty} dV_{z} g_{z}^{2} \Phi_{+,st}({\bm
x},Z,V_{z}) \delta(Z-z).
\end{eqnarray}
This expression can be rewritten in a more familiar form. First,
decompose it as
\begin{eqnarray}
\label{ap2.2} F_{z} & = & m \int d{\bm x} \int_{0}^{\infty} dZ
\int_{-\infty}^{\infty}dV_{z}\,   v_{z} g_{z}
\Phi_{+,st}({\bm x},Z,V_{z}) \delta(Z-z) \nonumber \\
&& - m \int d{\bm x} \int_{0}^{\infty} dZ \int_{-\infty}^{\infty}
dV_{z}\, \left[ v_{z} - \frac{M}{m+M}(1+\alpha_{P})g_{z} \right]
g_{z}
\Phi_{+,st}({\bm x},Z,V_{z}) \delta(Z-z).  \nonumber \\
\end{eqnarray}
Next, taking into account Eq.\ (\ref{3.2}), the second term on the
right hand side of this equality is seen to be equivalent to
\begin{eqnarray}
\label{ap2.3} && - m \int d{\bm x} \int_{0}^{\infty} dZ
\int_{-\infty}^{\infty} dV_{z}\,  v^{\prime}_{z}  g_{z}
\Phi_{+,st}({\bm x},Z,V_{z}) \delta(Z-z) \nonumber \\
&& = - m \int d{\bm x}^{*} \int_{0}^{\infty} dZ
\int_{-\infty}^{\infty} dV_{z}^{*}\,  v_{z}  g_{z}^{*}
\Phi_{+,st}({\bm
r},{\bm v}^{*},Z,V_{z}^{*}) \delta(Z-z) \nonumber \\
&&=  m \int d{\bm x} \int_{0}^{\infty} dZ \int_{-\infty}^{\infty}
dV_{z}\,  v_{z}  g_{z} \Phi_{-,st}({\bm x},Z,V_{z}) \delta(Z-z).
\end{eqnarray}
In the above transformations, ${\bm x}^{*} \equiv \{ {\bm r}, {\bm
v}^{*} \}$ and the exact boundary condition in Eq.\ (\ref{3.20}) has
been employed. Substitution of Eq. (\ref{ap2.3}) into Eq.
(\ref{ap2.2}) and use of the parity of $\Phi_{0,st}$ with respect to
$V_{z}$ as well as its independence from ${\bm r}_{\perp} $ gives
\begin{eqnarray}
\label{ap2.4} F_{z}& = & m \int d{\bm x} \int_{0}^{\infty} dZ
\int_{-\infty}^{\infty} dV_{z}\, v_{z} g_{z} \Phi_{0,st} ({\bm
x},Z,V_{z}) \delta(Z-z) \nonumber \\
&=& W n_{L} T_{L,z},
\end{eqnarray}
where $n_{L}$ is the density of the granular gas next to the piston
and $T_{L,z}$ a temperature parameter defined from the $z$-component
of the velocity. Its expression is given in Eq.\ (\ref{4.3}).

It is worth to stress that no assumption has been made over the
properties of $\Phi_{0,st}$ other than the associated with the
general symmetry properties of the steady macroscopic state under
consideration and the boundary condition (\ref{3.20}).

\section{The boundary condition at the vibrating wall}
\label{ap3}

Accordingly with the description of the vibrating wall located at
the bottom of the system given at the beginning of Sec.\ \ref{s3},
when a particle with velocity ${\bm v}$, being $v_{z}<0$, collides
with this wall, its velocity is instantaneously changed into ${\bm
v}^{\prime}$ defined by
\begin{equation}
\label{ap3.1} v^{\prime}_{z}= 2 v_{W}-v_{z}, \quad {\bm
v}_{\perp}^{\prime}={\bm v}_{\perp},
\end{equation}
Here, as already mentioned, $v_{W}$ is the (upwards) velocity of the
wall and ${\bm v}_{\perp}$ is the vector component of ${\bm v}$
perpendicular to the $z$-axis. The kinetic energy gained by the
particle in such a collision is
\begin{equation}
\label{ap3.2} \Delta \epsilon = 2m (v_{W}^{2}-v_{W}v_{z}).
\end{equation}
For physical initial conditions in which there are no particles
below the vibrating wall, the one-particle distributions function of
the gas at arbitrary later times verifies
\begin{equation}
\label{ap3.3} f({\bm x},t)= \Theta (z) f_{0}({\bm x},t),
\end{equation}
where $f_{0}({\bm x},t)$ and its derivatives can be taken regular at
$z=0$. Next, the above distribution function is decomposed at the
wall as
\begin{equation}
\label{ap3.4} f_{0}({\bm x},t) \delta(z) = f_{0}^{(+)} ({\bm x},t)
\delta (z) + f_{0}^{(-)}({\bm x},t) \delta (z),
\end{equation}
with
\begin{equation}
\label{ap3.5} f_{0}^{(\pm)}({\bm x},t) = \Theta (\pm v_{z})
f_{0}({\bm x},t).
\end{equation}

Conservation of the flux of particles at this wall is expressed by
\begin{equation}
\label{ap3.6} f_{0}^{(-)}({\bm x},t) |v_{z}| \delta(z) d{\bm v} =
f_{0}^{(+)} ({\bm x}^{\prime},t) |v^{\prime}_{z}| \delta (z) d {\bm
v}^{\prime},
\end{equation}
where ${\bm x}^{\prime} \equiv \{ {\bm r},{\bm v}^{\prime} \}$.
Using Eqs. (\ref{ap3.1}), the above relation can be transformed into
\begin{equation}
\label{ap3.7} f_{0}^{(+)}({\bm x},t) v_{z} \delta(z) = f_{0}^{(-)}
({\bm x}^{\prime},t) |v_{z}^{\prime}| \delta (z).
\end{equation}

The heat flux at the vibrating wall, $Q_{0}$, is given by the rate
of energy input per unit of section of the wall, i.e.
\begin{eqnarray}
\label{ap3.8} Q_{0}({\bm r}_{\perp},t) & = &\int dz \int d{\bm v}\
|v_{z}| \Delta \epsilon f_{0}^{(-)} ({\bm x},t ) \delta (z)
\nonumber \\
&=& \int dz \int {\bm v}\,  m v_{z}^{2} v_{W} f_{0}^{(-)} ({\bm
x},t) \delta (z)
\delta (z) \nonumber \\
& & + \int dz \int d{\bm v}\, m |v_{z}| v_{W} |2v_{W}-v_{z}|
f_{0}^{(-)} ({\bm x},t) \delta (z).
\end{eqnarray}
Consider the second term on the right hand side of the above
expression. Use of Eq.\ (\ref{ap3.7}) yields
\begin{equation}
\label{ap3.9} \int dz \int d{\bm v}\, m |v_{z}| v_{W} |2v_{W}-v_{z}|
f_{0}^{(-)} ({\bm x},t) \delta (z)= \int dz \int d{\bm v}\, m
v_{z}^{2}  v_{W} f_{0}^{(+)} ({\bm x},t) \delta (z),
\end{equation}
and substitution of this into Eq. (\ref{ap3.8}) gives
\begin{equation}
\label{ap3.10} Q_{0}({\bm r}_{\perp},t) = \int dz \int d{\bm v}\, m
v_{z}^{2} v_{W} f_{0}({\bm x},t) \delta (z) = v_{W} n_{0}({\bm
r}_{\perp},t) T_{0,z} ({\bm r}_{\perp},t),
\end{equation}
where
\begin{equation}
\label{ap3.11} n_{0}({\bm r}_{\perp},t) \equiv \int dz \int d{\bm
v}\, f_{0}({\bm x},t) \delta (z)
\end{equation}
and
\begin{equation}
\label{ap3.12} n_{0}({\bm r}_{\perp},t) T_{0,z} ({\bm r}_{\perp},t)
\equiv \int dz \int d{\bm v}\, m v_{z}^{2} f_{0}({\bm x},t) \delta
(z).
\end{equation}
Equation (\ref{4.23}) in the main text follows by particularizing
Eq. (\ref{ap3.10}) for the one-dimensional steady state considered
in this paper and assuming isotropy of the pressure tensor of the
gas next to the vibrating wall.

The issue of the energy injected in a granular gas by means of an
elastic vibrating wall moving in a sawtooth way has been addressed
previously in ref. \cite{BRyM00}. The arguments there are restricted
to a particular state with the gas modeled by means of Gaussian
distributions. Moreover, although the results reported in
\cite{BRyM00} are consistent with Eq.\ (\ref{ap3.10}), the
simplicity of this latter form of expressing the result was not
realized there.

\end{document}